\documentclass[a4paper]{article}

\usepackage[colorlinks=true,dvips]{hyperref}
\usepackage{amsmath,amsfonts,amssymb}
\usepackage{float}
\usepackage{graphicx}
\usepackage{listings}
\usepackage{xcolor}

\title{A loopless and branchless $O(1)$ algorithm to generate the next Dyck word.%
\footnote{%
  This work is licensed under a
  \href{http://creativecommons.org/licenses/by-sa/4.0/}
  {Creative Commons Attribution-ShareAlike 4.0 International License}.
}}
\author{Cassio Neri \\}
\date{19 July 2014%
  \footnote{First revision uploaded to \cite{Neri} on 22 Jul 2014 and current revision on 15 Feb 2018.}
}

\hypersetup{  
  pdfinfo={  
    Title={A loopless and branchless O(1) algorithm to generate the next Dyck word.},  
    Author={Cassio Neri},
    Subject={Algorithm},
    Keywords={Dyck words, Balanced parentheses, Monotonic paths}
  }  
}

\begin{document}

\maketitle

\begin{abstract}

Let \verb!integer! be any C/C++ unsigned integer type up to $64$-bits long.
Given a Dyck word the following code returns the next Dyck word of the same size, provided it exists.
\\
\begin{lstlisting}[caption={Bitwise tricks to generate a Dyck word.},label=bitwise]
integer next_dyck_word(integer w) {
  integer const a = w & -w;
  integer const b = w + a;
  integer       c = w ^ b;
                c = (c / a >> 2) + 1;
                c = ((c * c - 1) & 0xaaaaaaaaaaaaaaaa) | b;
  return c;
}
\end{lstlisting}
\end{abstract}

\section{Introduction}

A $2n$-bits Dyck word is a string containing exactly $n$ 1s and $n$ 0s and such that each of its prefix substrings contains no more 0s than 1s.

\begin{figure}[ht]
  \def\svgwidth{\textwidth}
  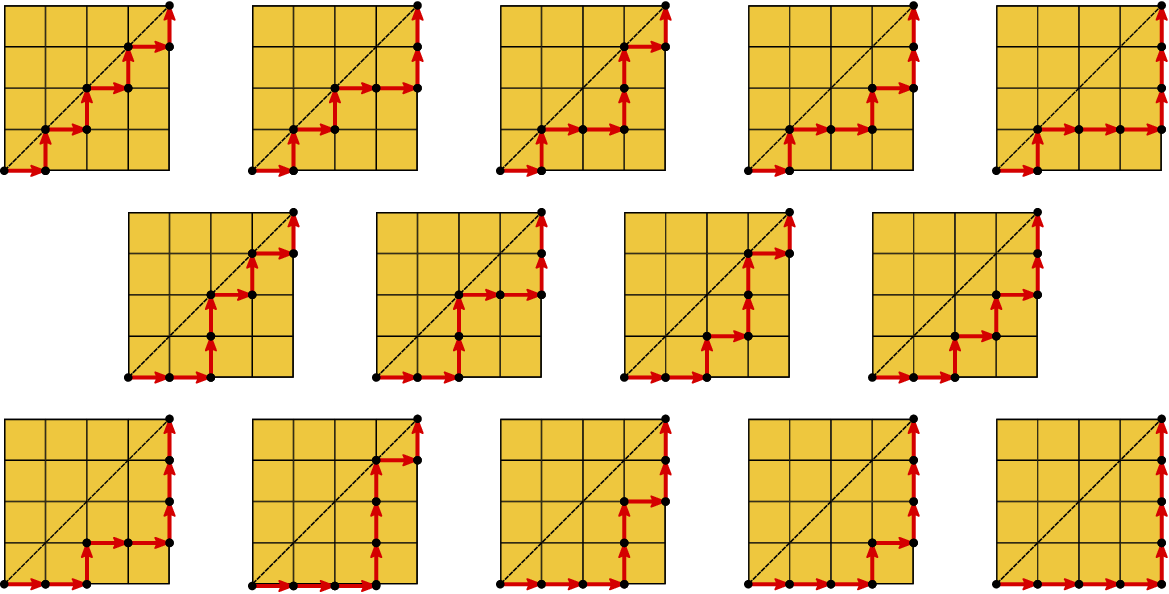
  \hfill\
  \caption{Monotonic paths on a $4\times 4$ grid starting at the lower left corner, finishing at the upper right corner and staying below diagonal.
  There are 14 of them.
  The first path, for instance, corresponds to the Dyck word $10101010$ and the fith one to $10111000$.}
  \label{fig:Paths-4x4}
\end{figure}

Dyck words appear in a vast number of problems \cite{Wikipedia-Catalan}.
Consequently, generating them has many applications.
For instance, if 1s and 0s are replaced with openning and closing parentheses, then a Dyck word is a combination of $n$ properly balanced pairs of parentheses.
When 1 denotes a move rightwards and 0 denotes a move upwards, a Dyck word represents a monotonic path along the edges of an $n\times n$ grid that starts at the lower left corner, finishes at the upper right corner and stays below diagonal.
Figure \ref{fig:Paths-4x4} shows all $14$ such paths on a $4\times 4$ grid.

For the sake of clarity, another implementation of essentially the same algorithm is presented.
This one works on strings of two different arbitrarily chosen symbols (e.g., opening and closing parentheses).
Contrarily to the implementation in Listing \ref{bitwise}, the second one has explicit loops and branches and has $O(n)$ time complexity.

The apparent contradiction between being two implementations of ``essentially the same'' algorithm but one being loopless, brancheless and $O(1)$ and the other having loops, branches and being $O(n)$ can be explanained as follows.

To perform bitwise and arithmetic operations the hardware somehow ``runs" loops on its transistors but, as far as I understand, for the sake of software complexity, these operations are considered as being loopless, branchless and $O(1)$.
In that sense we can categorize the algorithm in Listing \ref{bitwise} as loopless, branchless and $O(1)$.
The second implementation (through software code) simply explicits the loops and branches ``ran" by the hardware when Listing \ref{bitwise} is executed.

Both implementations are $O(1)$ in space.

{\bf Disclaimer}: I am not a Computer Scientist and I am not aware of the state of the art.
I do not claim the algorithm in Listing \ref{bitwise} has not been discovered/invented before but I believe there is a strict positive probability that it has not.
I have performed a not very thorought search on the net and I have failed to find anything similar.
The closest I know is the Gosper's hack \cite{Wikipedia-Combinatorial} which, I must say, was an inspiration of Listing \ref{bitwise}.
Furthermore, this implementation borrows its first operations from Gosper's hack.

{\bf Update 1}: Following extra search, I came across exercise 23 of Kunth \cite{Knuth}:

\begin{quote}
A sequence of nested parentheses can be represented as a binary number by putting a 1 in the position of each right parenthesis.
For example, '(())()' corresponds in this way to $(001101)_2$, the number $13$.
Call such a number a {\em parenthesis trace}.
\begin{quote}
a) What are the smallest and largest parenthesis traces that have exactly $m$ 1s? \\
b) Suppose $x$ is a parenthesis trace and $y$ is the next parenthesis trace with the same number of 1s.
Show that $y$ can be computed from $x$ with a short chain of operations analogous to Gosper's hack.
\end{quote}
\end{quote}
Listing \ref{bitwise} is, basically, a solution for the exercise except that it reverses the interpretation of bits.
Due to this reversal, the solution presented in \cite{Knuth} is differ from Listing \ref{bitwise} and uses a, potentially costly, \verb!sqrt! operation.

{\bf Update 2}: I also came across Warren \cite{Warren} which explains Gosper's hack and suggests the optimizations based on \verb!popcount! or \verb!ctz! that I have added to this article in February 2018.
It also made me realize that Listing \ref{bitwise} is closer to Gosper's hack than I have initially thought.

\section{Definitions}

Let $n\in\mathbb{N}$ and $B = \{1, 0\}$.
A $2n$-bits {\bf word} is an element $w = (w_1, ..., w_{2n})$ of $B^{2n}$.
To easy notation, we most often drop parentheses and commas.
(For instance, $(1, 0, 1, 0, 1, 0, 1, 0)$ becomes $10101010$.)
Finally, by abuse of notation, we identify $w = (w_1, ..., w_{2n})$ with the number whose binary expansion is $w$. More precisely, with
\[
\sum_{i = 0}^{2n - 1} w_{2n - i} 2^i.
\]
Thanks to this identification we can order words, talk about minimum, maximum, etc.

For $w\in B^{2n}$ and $i\in\{1, ..., 2n\}$ we define
\begin{align*}
N_1(w, i) & := \#\{ j \le i\ ;\ w_j = 1 \}, \\
N_0(w, i) & := \#\{ j \le i\ ;\ w_j = 0 \}.
\end{align*}
In plain English, $N_1(w, i)$ is the number of 1s in $w$ appearing before or at $i$-th position
and $N_0$ is the number of 0s before or at position $i$.

We say that $w\in B^{2n}$ is a $2n$-bits {\bf Dyck word} (or simply a Dyck word when $n$ is implicit) if
\begin{equation}
\label{eq:NoCross}
N_1(w, i) \ge N_0(w, i) \quad \forall i\in\{0, ..., 2n - 1\},
\end{equation}
and
\begin{equation}
\label{eq:Path}
N_1(w, 2n) = N_0(w, 2n) = n.
\end{equation}
In plain English, before or at the $i$-th position, the number of 1s must be no lesser than the number
of 0s.
In adition the total numbers of 1s and 0s match.

For the mathematically trained eye, property \eqref{eq:NoCross} becomes easier to spot%
\footnote{At least for small values of $n$ or not so small if you are a Lisp programmer.}
when 1s and 0s are replaced, respectively, with openning and closing parentheses.
In this case, the Dyck word contains $n$ pairs of correctly matched open and close parentheses.
For instance, $10101010$ becomes ()()()() and $10111000$ becomes ()((())).

\section{The minimum and the maximum Dyck word}

The function in Listing \ref{bitwise} generates the succesor of a given Dyck word.
Therefore, to kick off and stop generating all words, we need to know the minimum and maximum Dyck word of a ginven size.
(Recall that ``minimum'' and ``maximum'' refer to the order of integer numbers.)

We claim that the minimum $2n$-bits Dyck word is
\[
\breve w := \underbrace{10\cdots10}_{n\text{ pairs}}
\]
(or $()\cdots()$ in the parenthetical representation).
More precisely, $\breve w = (\breve w_1, ..., \breve w_{2n})\in B^{2n}$ is given by $\breve w_i = 1$, if $i$ is odd, and $\breve w_i = 0$, if $i$ is even.

Before proving our claim, we compute $\breve w$:
\begin{align*}
\breve w &= 2^{2n - 1} + 2^{2n - 3} + \cdots + 2^3 + 2^1 \\
&= 2 \cdot 2^{2n - 2} + 2\cdot 2^{2n - 4} + \cdots + 2\cdot 2^2 + 2\cdot 2^0 \\
&= 2\cdot 4^{n - 1} + 2\cdot 4^{n - 2} + \cdots + 2\cdot 4 + 2\cdot 4^0 = 2\sum_{i = 0}^{n - 1}4^i = \frac23(4^n - 1).
\end{align*}
Calculating this number on real computers needs care to avoid overflow.
For instance, if $n = 32$ then $4^n = 4^{32} = 2^{64}$ which is one more than the maximum natural number representable by a $64$-bits unsigned integer type.
In practice, however, this is a ``minor'' issue because generating all $64$-bits Dyck words would take ``forever'' given that the number of $2n$-bits Dyck words grows factorially \cite{Wikipedia-Catalan} with $n$.

We shall now prove that $\breve w$ is the minimum Dyck word.
Suppose by contradiction that there's a Dyck word $w = (w_1, ..., w_{2n})$ such that $w < \breve w$.
In particular, $w \ne \breve w$ and let $i$ be the minimum index such that $w_i \ne \breve w_i$.
Because all Dyck words start with 1 we must have $i > 1$ and since $w < \breve w$, we have $w_i = 0$ and $\breve w_i = 1$.
By construction of $\breve w$, it follows that $i$ is odd, $i - 1$ is even and $N_1(\breve w, i - 1) = N_0(\breve w, i - 1)$.
The same holds for $w$ because it shares the first $i - 1$ bits with $\breve w$.
Now, $w_i = 0$ and thus  $N_0(w, i) = N_0(w, i - 1) + 1$ and $N_1(w, i) = N_1(w, i - 1)$ which yields $N_0(w, i) = N_1(w, i) + 1$, contradicting \eqref{eq:NoCross}.

Its much easier to see that the maximum $2n$-bits Dyck word is
\[
\hat w := \underbrace{1\cdots1}_{n\text{ times}}\ \underbrace{0\cdots0}_{n\text{ times}},
\]
which values $2^{2n} - 2^n$.

\section{The next Dyck word}

Let $w = (w_{2n - 1}, ..., w_0)$ be a Dyck word.
In this section we characterize the Dyck word that succeeds $w$, that is, the smallest Dyck word of the same size which is greater than $w$.

Assuming that $w\ne\hat w$, {\it i.e.}, $w$ is not the maximum Dyck word, there exists at least one index $i \in\{1, ..., 2n\}$ such that $w_i = 0$ and $w_{i + 1} = 1$.
Let $k$ be the maximum of such indices.
Since a Dyck word cannot start with a 0 or finish with a 1, we have $1 < k < k + 1 < 2n$.
By the maximality of $k$, after this position there is no 0 followed by 1.
More precisely, there's a (possibly empty) sequence of 1s followed by a non empty sequence of 0s up to the end.
Hence, $w$ has this form:
\[
w = (\underbrace{w_1, ..., w_{k - 1}}_{\text{prefix}}, \underbrace{0}_{w_k}, \underbrace{1}_{w_{k + 1}}, \underbrace{1, ..., 1}_{x\text{ times}}, \underbrace{0, ..., 0}_{y\text{ times}}),
\]
with $x\ge 0$ and $y > 0$.
We shall prove that the successor $\tilde w$ of $w$ has this form:
\[
\tilde w = (\underbrace{w_1, ..., w_{k - 1}}_{\text{prefix}}, \underbrace{1}_{\tilde w_{k}}, \underbrace{0}_{\tilde w_{k + 1}}, \underbrace{0, ..., 0}_{y - x\text{ times}}, \underbrace{1, 0, ..., 1, 0}_{x\text{ pairs of 1, 0}}).
\]

Notice that the prefixes of $w$ and $\tilde w$ are the same.

Before seeing the formal proof, we present the idea which is quite simple.
Since we want to increase a binary number the least as possible, we must flip a bit 0 into a bit 1 and this bit must be as much to the right as possible.
None of the $y$ bits 0 ending $w$ can be flipped without breaking \eqref{eq:NoCross}.
The first opportunity is $w_k$.
Following $\tilde w_k$, we want $\tilde w$ to be as least as possible and hence fill it up with 0s as much as we can without breaking \eqref{eq:NoCross}.
This will leave us with a $(2n - 2x)$-bits Dyck word on the left hand side of $\tilde w$ and to complete a $2n$-bits Dyck word we need to append the minimum $2x$-bits Dyck word.

First we shal show that $\tilde w$ is well defined, that is, $y - x \ge 0$, the size of $\tilde w$ is $2n$ and $\tilde w$ verifies the properties \eqref{eq:NoCross} and \eqref{eq:Path}.

The total number of 1s in $w$ is $N_1(w, k - 1) + 1 + x$ and the total number of 0s is $N_0(w, k - 1) + 1 + y$.
Since $w$ is a Dyck word, these numbers match and we obtain $N_1(w, k - 1) + 1 + x = N_0(w, k - 1) + 1 + y$.
Hence,
\begin{equation}
\label{eq:Balance}
N_1(w, k - 1) + 1 = N_0(w, k - 1) + 1 + y - x \Longrightarrow y - x = N_1(w, k - 1) - N_0(w, k - 1).
\end{equation}
Again, because $w$ is a Dyck word, $N_1(w, k - 1) - N_0(w, k - 1)$ is positive and so is $y - x$.

The last two segments of $w$ have total size $x + y$ and the last two segments of $\tilde w$ have total size $y - x + 2 x = x + y$.
Hence $\tilde w$ is also $2n$-bits long and $k + 1 + y - x = 2(n - x)$.

Because of the common prefix with $w$, $\tilde w$ verifies property \eqref{eq:NoCross} for any $i \le k - 1$.
Obviously, it also verifies \eqref{eq:NoCross} for $i = k$ ($\tilde w$ gets an extra $1$).

Notice that for any $i\in\{k + 1, ..., 2(n - x)\}$, $\tilde w_i = 0$.
Hence, if $\tilde w$ fails to verify \eqref{eq:NoCross} for any $i$ in this set of indices, then it fails to verify \eqref{eq:NoCross} for $i = 2(n - x)$.
However, $N_1(\tilde w, 2(n - x)) = N_1(w, k - 1) + 1$ and $N_0(\tilde w, 2(n - x)) = N_0(w, k - 1) + 1 + y - x$ and from \eqref{eq:Balance} we obtain that these numbers are equal.
This proves that \eqref{eq:NoCross} holds for $i \le 2(n - x)$.
Furthermore, we have proven that $N_1(\tilde w, 2(n - x)) = N_0(\tilde w, 2(n - x))$.

From the $2(n - x)$-th bit onwards, the sequence is alternating and it follows that
\begin{equation}
\label{eq:Alternating}
\text{if } i \ge 2(n - x) \text{ is even, then }\tilde w_i = 0 \text{ and } N_1(\tilde w, i) = N_0(\tilde w, i).
\end{equation}

To prove that $\tilde w$ is the smallest Dyck word which is greater than $w$, assume there's another Dyck word $v = (v_1, ..., v_{2n})$ such that $w < v < \tilde w$.
Well, obviously, $v$ must have the same prefix as $w$ and $\tilde w$.
What about $v_k$?
It is either 0 or 1 and we slipt in two cases.

If $v_k = 0 = w_k$ then, because $w < v$ and $w_i = 1$ for all $i\in\{k + 1, ..., k + 1 + x\}$, the common part between $v$ and $w$ must span up to index $k + 1 + x$.
For $i > k + 1 + x$, $w_i = 0$ and for $w < v$ to hold, at least one of the last $y$ bits of $v$ must be 1.
Then the number of 1s in $v$ is greater than the number of 1s in $w$ which violates the fact that any Dyck word has exactly $n$ 1s.

Similarly, if $v_k = 1 = \tilde w_k$, then because $v < \tilde w$ and $\tilde w_i = 0$ for $i\in\{k + 1, ..., 2(n - x)\}$ the common part between $v$ and $\tilde w$ must span up to index $2(n - x)$.
But, since $v\ne\tilde w$, there exists $i > 2(n - x)$ such that $v_j\ne\tilde w_j$.
Let $i$ be the minimum of such indices.
For $v < \tilde w$ to hold, it is necessary that $v_i = 0$ and $\tilde w_i = 1$.
From \eqref{eq:Alternating} it follows that $i$ is odd and $i - 1\ge 2(n - x)$ is even.
Again from \eqref{eq:Alternating} we obtain $N_1(\tilde w, i - 1) = N_0(\tilde w, i - 1)$ and the same holds for $v$ thanks to its common part with $\tilde w$.
But $v_i = 0$ and thus $N_1(v, i) = N_1(v, i - 1)$ and $N_0(v, i) = N_0(1, i - 1) + 1 = N_1(v, i - 1) + 1$ which violates \eqref{eq:NoCross}.

\section{Two implementations}

The C++ function in Listing \ref{string} (code available at \cite{Neri}) implements the algorithm presented in the previous section.
It does not allocate memory for the output and, instead, performs the manipulations in-place.
It makes no use of helper containers and, therefore, it's $O(1)$ on space.

The algorithm scans the word backwards up to a certain point.
Then it advances forward changing the bits up to the rightmost.
Hence, in the worse case, the program scans the whole word twice implying complexity $O(n)$ on time.

The function takes a Dyck word \verb+w+ made of \verb+one+s and \verb+zero+s (provided as arguments) and transforms it into the next Dyck word, if it exists, otherwise, it clears the word.
The behaviour is undefined if \verb+w+ is not a Dyck word of \verb+one+s and \verb+zero+s.

\begin{lstlisting}[float,caption={String manipulation to generate a Dyck word.},label=string]
void next_dyck_word(std::string& w, char const one, char const zero) {
  
  unsigned const m = w.size() - 1;
  unsigned       y = 0;
  unsigned       x = 0;

  for (unsigned i = m; i > 0; --i) {

    if (w[i] == zero)
      ++y; // Counter for trailing 0s.

    else if (w[i - 1] == zero) {
      
      // Found greatest i such that w[i] = zero and w[i + 1] = one.
      
      // Change these two chars.
      w[i - 1] = one;
      w[  i  ] = zero;
      
      // Overwrite the following next y - x chars to zero.
      for (y = y - x; y != 0; --y)
        w[++i] = zero;
      
      // Overwrite the remaining chars with alternating ones and zeros.
      while (i < m) {
        w[++i] = one;
        w[++i] = zero;
      }
      return;
    }

    else
      ++x; // Counter for 1s that precede the trailing zeros.
  }
  w.clear(); // Failed to produce a Dyck word, then clear w.
}
\end{lstlisting}

We shall consider now the implementation shown in Listing \ref{bitwise}.
The function shown there, similarly to the one in Listing \ref{string}, assumes that the input is a Dyck word.
If it is not, then this is a pre-condition violation which yields undefined behaviour.
In addition (and opposite to the implementation in Listing \ref{string}), another pre-condition is that the input is not the maximum Dyck word of its size.
Failing to verify this condition, again, produces undefined behaviour.


The abstract says that \verb!integer! is a C/C++ {\em unsigned} integer type and the first line in the function's body takes the opposite of \verb!w!!
This is intended and works as expected on C and C++ conforming implementations because on these systems unsigned integer types have $2^N$-modular arithmetics, where $N$ is the size in bits of the integer type.

Other platforms might have different unsigned integer types and, at this point, it is useful to list the properties the type must verify for the algorithm to work.

The first rule is obvious but worth saying: the type must implement the usual binary representation of unsigned integer numbers.
In particular, all values in the range $[0, 2^N - 1[$ are representable.
$2^N$-modular arithmetics is sufficient but not necessary and only the opposite of \verb!w! must be as per $2^N$-modular arithmetics (simply put, the opposite of $w$ is $2^N - w$).
If this is an issue, then the slightly slower variation for the first line can be used:
\begin{lstlisting}
  integer const a = w & (~w - 1);
\end{lstlisting}

For the other operations, usual arithmetic rules are enough because the (mathematical) results of additions, subtractions, multiplications and divisions stay in the $[0, 2^N - 1[$ range.
Moreover, when the division is performed, \verb!c! is a multiple of \verb!a! (see below) and, therefore, no truncation or division by 0 occurs.

We assume usual semantics also for the bitwise operators \verb!&!, \verb!^!, and \verb!|! (and \verb!~! if you use the alternative to \verb!-w! shown above).
Finally, the right shift rotation inserts 0s in the gaps on the left but if, this is not the case, we can replace the right shift of two bits by a division by four, provided that the division truncates the result.

The pleasure of verifying the details that the arithmetic and bitwise operations in Listing \ref{bitwise} reproduce the construction explained in previous section is left to the reader.
I shall provide an overall picture though.

Let $\tilde k$, $x$, $y$ and $\tilde w$ be as in the previous section.

The only 1-bit of \verb!w & -w! is the same as rightmost 1-bit of \verb!w! and is located at position $k + 1 + x$.
In other words, \verb!a! is the largest power of $2$ that divides \verb!w!, namely, $\verb!a! = 2^{2n - k - 1 - x} = 2^y$.

The bits of $\verb!b! = \verb!w + a!$ matches those of $\tilde w$ up to position $k + 1 + y - x = 2(n - x)$ and are 0s afterwards.
Hence, the first two lines of code do almost everything.
The following three will create the sequence of $x$ pairs of $1$ and $0$ (which we recognize as the minimum $2x$-bits Dyck word) and apply operator \verb!|! to it and \verb!b!.

The first value assigned to \verb!c! has this form
\[
(\underbrace{0, ..., 0}_{\text{prefix}}, \underbrace{1}_{w_k}, \underbrace{1}_{w_{k + 1}}, \underbrace{1, ..., 1}_{x\text{ times}}, \underbrace{0, ..., 0}_{y\text{ times}}),
\]
which values $(2^{x + 2} - 1)\cdot 2^y$.
Since $\verb!a! = 2^y$, dividing \verb!c! by \verb!a! has the effect of right-shifting \verb!c! by $y$ bits.
Shifting two extra bits produces the number whose $x$ least significant bits are 1, that is, $2^x - 1$.
After the addition to $1$, \verb!c! gets the value $2^x$.
Squaring produces $2^{2x}$ and decrementing yields $2^{2x} - 1$ which is a mask for the $2x$ rightmost bits.

Now the magic number comes in (at this point, you have probably guessed what this is).
It is the $64$-bits number whose binary expansion is an alternating sequence of 1s and 0s or, in other terms, it is the minimum $64$-bits Dyck word.

Applying operator \verb!&! to the mask and the magic number produces the $2x$-bits minimum Dyck word to fill the gap in \verb!b!.

\section{Getting faster}
\label{Sec:Getting}

The code in Listing \ref{bitwise} is very efficient but programmers who love to code close to the metal might raise the question about the ``expensive'' division.
This operation can, indeed, be removed.

As explained in the previous paragraph, after \verb!b! is set what is missing is applying operator \verb!|! to \verb!b! and the minimum $2x$-bits Dyck word.
Computing the minimum Dyck word of a given size is straightforward provided that we know the size but, here, we do not.
We have theoretically called it $2x$ but we do not know its value.
What we do know is that \verb!c! has $2x + 2$ bits 1.
Provided that we have a fast way to count bits, {\it i.e.}, a \verb!popcount! function, then we can compute the value of $x$ and, subsequently, the minimum $2x$-bits Dyck word.
This underlyies this new implementation:
\\
\begin{lstlisting}
integer next(integer w) noexcept {
  integer a = w & -w;
  integer b = w + a;
  integer       c = w ^ b;
  unsigned      x = popcount(c) - 2;
                c = 0xaaaaaaaaaaaaaaaa >> (32 - x) >> (32 - x);
  return b | c;
}
\end{lstlisting}

Notice that we shift twice by $32 - x$ instead of once by $64 - 2x$.
The reason is that $x$ can be $0$ and right shifting a $64$-bits unsigned integer by $64$ bits yields undefined behaviour.
Besides, calculating $2x$ is most efficiently done by a left shift $\verb!x! << 1$ hence there is no performance advantage of ``\verb!>> (64 - (x << 1))!'' over ``\verb!>> (32 - x) >> (32 - x)!''.

Modern hardwares provide a \verb!popcount! instruction and some compilers expose it through intrinsic functions.
On these platforms the implementation above can be used.

Older CPUs do not provide \verb!popcount! but implement a \verb!ctz! instruction to count the number of trailing 0s of a binary number.
For the specifc bit pattern of \verb!c! we can use \verb!ctz! to work around the lack of \verb!popcount!.
Indeed, recall that when we first set \verb!c! 
\[
\verb!c! = (2^{x + 2} - 1)\cdot 2^y = 2^{x + 2 + y} - 2^y = 2^{x + 2 + y} - \verb!a!.
\]
Hence, $\verb!c! + \verb!a! = 2^{x + 2 + y}$.
Moreover,
\[
\verb!popcount(c)! = x + 2 = (x + 2 + y) - y = \verb!ctz(!2^{x + 2 + y}\verb!)! - \verb!ctz(!2^y\verb!)! = \verb!ctz(c + a)! - \verb!ctz(a)!.
\]

For CPUs that are even older and no \verb!ctz! is provided, one can sill use a software implementation \cite{Wikipedia-Popcount} of \verb!popcount! which might outperform the implementation in Listing \ref{bitwise}.

\section{Acknowledgements}

I thank my friend Prof. Lorenz Schneider who introduced me to the Gosper's hack around 2011.
I am also grateful towards {\em mosh111}, whoever he/she is, who introduced me to the problem of Dyck word generation through his/her post in \cite{CareerCup} on 16 July 2014.

\end{document}